\documentclass[epj]{webofc}
\usepackage[varg]{txfonts}   
\usepackage{hyperref}
\usepackage{wrapfig}
\newcommand{\fullwidth}{0.99\textwidth}
\newcommand{\figwidth}{0.9\textwidth}
\newcommand{\smallhalffigwidth}{0.5\textwidth}
\newcommand{\verysmallhalffigwidth}{0.35\textwidth}

\newcommand{\tablewidth}{0.65\textwidth}
\newcommand{\VEC}[1]{{\boldsymbol{#1}}}
\newcommand{\MAT}[1]{{\boldsymbol{#1}}}
\wocname{EPJ Web of Conferences}
\woctitle{CONF12}
\begin{document}
\selectlanguage{english}
\title{Data Unfolding Methods in High Energy Physics}
%
%

\author{Stefan Schmitt\inst{1}\fnsep\thanks{\email{sschmitt@mail.desy.de}}}

\institute{DESY, Notkestra\ss{}e 85, Hamburg, Germany}

\abstract{%
A selection of unfolding methods commonly used in High Energy Physics
is compared. The methods discussed here are: bin-by-bin correction factors, 
matrix inversion,
template fit, Tikhonov regularisation and two examples of iterative
methods. Two procedures to choose the strength of the regularisation
are tested, namely the L-curve scan and a scan of global correlation
coefficients. The advantages and disadvantages of the unfolding
methods and choices of the regularisation strength are discussed using
a toy example.
}
\maketitle

\section{Introduction}
\label{intro}
In high energy  physics, typical measurements are based on counting
experiments. Events are detected and later classified depending on
their properties. Cross sections, for example, are determined
from event counts, where the event properties are restricted to 
certain regions in phase space (bins), divided by the integrated luminosity.
The observed event counts are different from the expectation for an
ideal detector mainly because of three effects:
\begin{description}
\item[Detector effects:] the event properties such as energy or
  scattering angle are measured only with finite precision and limited
  efficiency. Events may be reconstructed in the wrong bin or may get lost.
\item[Statistical fluctuations:] the observed number of events is drawn
  from a Poisson distribution. The measurement provides an estimate of
  the Poisson parameter $\mu$. Commonly, the square root of the
  number of event counts is assigned as ``statistical uncertainty''.
\item[Background:] events similar to the signal may also be produced
  by other processes.
\end{description} 
The process of extracting information about the truth content of the
measurement bins, given the observed measurements, is referred to as
``unfolding''. In mathematics, the general problem is formulated as an
integral equation of the type
\begin{eqnarray}
 \int k(y,x)f(x)dx & = & g(y)\,.
\label{eqn:integral}
\end{eqnarray}
Given the observations $g(y)$ and the kernel $k(y,x)$ one seeks to
know the function $f(x)$. It is well-known that for this type of
equation small changes of $g(y)$ may result in large changes of $f(x)$.

In the following, a simpler version of equation
\ref{eqn:integral}, corresponding to a finite number of bins, is
studied. The distribution $f(x)$ is replaced by a vector $\VEC{x}$ of
dimension $M_X$, where the components $x_j$ correspond to the 
expected number of events in a bin $j$ at ``truth level''. 
Similarly, the function $g(y)$ is replaced by a vector $\VEC{\mu}$ of
dimension $M_y$ and its components $\mu_i$ correspond to the expected
number of events on ``detector level''. The two vectors are connected
by the folding equation
\begin{eqnarray}
 \MAT{A}\VEC{x} +\VEC{b} & = & \VEC{\mu}\,,
\end{eqnarray}
where the elements $A_{ij}$ of the response matrix $\MAT{A}$ specify
the probability 
to find an event produced in bin $j$ to be measured in bin $i$. 
The expected number of background events is described by the vector $\VEC{b}$.
The matrix $\MAT{A}$ and the vector $\VEC{b}$ are assumed to be known. In a real
experiment, these numbers may have limited precision, leading to
systematic uncertainties.
In many cases the $A_{ij}$ are estimated using Monte
Carlo techniques to simulate the signal process and detector effects,
\begin{eqnarray}
 A_{ij} & = & \frac{N_{ij}^{\text{MC,rec}\wedge\text{gen}}}{N_{j}^{\text{MC,gen}}}\,.
\label{eqn:folding}
\end{eqnarray}
The number $N_{ij}^{\text{MC,rec}\wedge\text{gen}}$ corresponds to the number of Monte
Carlo events generated in truth bin $j$ and reconstructed on the
detector in bin $i$. The number $N_{j}^{\text{MC,gen}}$ is the total
number of Monte Carlo events generated in truth bin $j$, including
events which are not reconstructed in any of the bins $i$.
The reconstruction efficiency is given by
$\epsilon_j=\sum_i A_{ij}$. Events which are reconstructed in a bin
$j$ but are generated outside any of the $M_X$ generator bins are
attributed to the background $b_j$.

As the experiment is performed, numbers $y_i$ are observed
instead of the expectation value $\mu_i$. The differences of the
vector of observations 
$\VEC{y}$ and the expectation $\VEC{\mu}$ are amplified in the
unfolding process.
For counting experiments, the integer event counts $y_i$ are drawn
from a Poisson distribution,
$P(y_i;\mu_i)=\exp(-\mu_i)(\mu_i)^{y_i}/y_i!$~. In the large sample
limit, the event counts $y_i$ are taken to follow multivariate
Gaussian distributions, with mean $\mu_i$ and a fixed covariance matrix
$\MAT{V_{yy}}$. The covariance matrix is diagonal in case of
statistically independent bins. The diagonal elements often are approximated
using the observations $y_i$ as the variances. 

The result of the unfolding process is an estimator $\VEC{\hat{x}}$ of
the truth distribution $\VEC{x}$ and a corresponding covariance matrix
$\MAT{V_{xx}}$. The uncertainties $\delta_j$ and 
correlation coefficients $\rho_{jk}$ of two bins $j$ and $k$ are given by
\begin{eqnarray}
\delta_j = \sqrt{(\MAT{V_{xx}})_{jj}}, & \quad \text{and} \quad &
\rho_{jk} = \frac{(\MAT{V_{xx}})_{jk}}{\delta_j \delta_k}\,.
\end{eqnarray}
The global correlation coefficient of bin $j$ is defined as
\begin{eqnarray}
\rho_j &=& \sqrt{1-\left(\left(\MAT{V_{xx}}\right)_{jj} \left(\MAT{V_{xx}}^{-1}\right)_{jj}\right)^{-1}}\,.
\end{eqnarray}

In this paper, a few selected unfolding
algorithms are discussed together with methods to verify the unfolding
procedures and to choose parameters of the algorithms. Unfolding
algorithms and their application in high-energy physics and elsewhere
are also widely discussed in dedicated workshops,
e.g.~\cite{phystat2011} and in literature, e.g.~\cite{kaipio2007,hansen2010}.

\section{Toy example}
A simple toy example is used here to test and compare various unfolding
algorithms. It is included in the TUnfold package
\cite{tunfold}, example number 7.
\begin{figure}[h]
\centering
\includegraphics[width=\figwidth,clip]{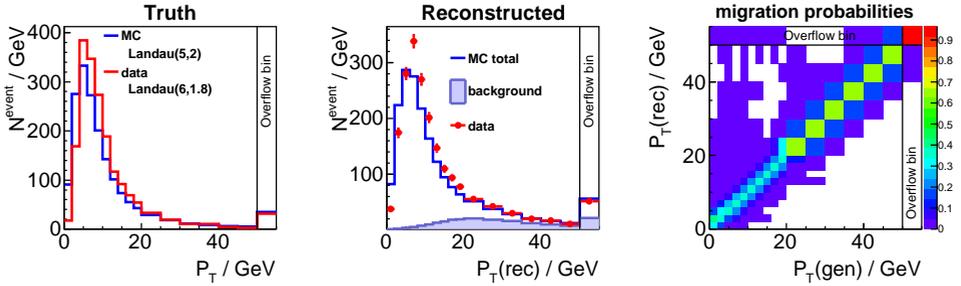}
\caption{Toy example, distribution on truth level (left), on
  reconstructed level (middle), response matrix (right).}
\label{fig:exampleTRA}       
\end{figure}
A heavy
particle is produced with a given transverse momentum $P_T$
distribution and decays into two massless particles. The energy and
angles of the decay products in the laboratory frame are smeared by
resolution functions.
The observed $P_T$ is calculated from the vector sum of the
two reconstructed particles. Background is also generated and
contributes in this example mainly at large $P_T$.
For the $P_T$ distribution on truth level, Landau
distributions are used. The corresponding parameters differ between the ``data''
and the ``simulated'' events, where the latter are used to
construct the response matrix. The resulting $P_T$
distributions are shown in Fig.~\ref{fig:exampleTRA} on truth and 
detector level. The differences between the two parameterisations are
clearly visible, both on truth and on reconstructed level. The
response matrix indicates a moderate detector resolution at low $P_T$,
where a fine binning is used.

\section{Testing unfolding results}
Given a method to estimate $\VEC{\hat{x}}$ and the covariance
$\MAT{V_{xx}}$ for a given vector of 
observations $\VEC{y}$, it is desirable to judge on the quality of this
estimator. Two classes of tests are defined here, ``Data tests'' and
``Closure tests''.

\subsection{Data tests}
The folding equation \ref{eqn:folding} can be applied to the
unfolding result, i.e.~one may compare 
$\MAT{A}\VEC{\hat{x}}+\VEC{b}$ with the observation $\VEC{y}$. The most
basic comparison is to verify the normalisation by calculating
\begin{eqnarray}
Y_{\text{unf}} := 
\sum_{i=1}^M\left(\MAT{A}\VEC{\hat{x}}+\VEC{b}\right)_i\quad &
\text{and} & \quad Y_{\text{data}} :=  \sum_{i=1}^M y_i\,.
\end{eqnarray}
The expectation is to find $Y_{\text{unf}}=Y_{\text{data}}$.
Another test is to calculate a $\chi^2$ sum,
\begin{eqnarray}
\chi^2_{A}=(\MAT{A}\VEC{\hat{x}}+\VEC{b}-\VEC{y})^{\text{T}}(\MAT{V_{yy}})^{-1}
(\MAT{A}\VEC{\hat{x}}+\VEC{b}-\VEC{y})\,.
\label{eqn:chi2obs}
\end{eqnarray}
In the large sample limit, one expects to find $\chi^2_A$ distributed
with $M_y-M_x$ degrees of freedom, unless a strong level of regularisation
is introduced by the unfolding procedure. In particular, for the case
$M_y=M_x$, the $\chi^2$ sum is expected to be zero and hence 
$Y_{\text{unf}}=Y_{\text{data}}$.
For $M_y>M_x$, the
quantiles of the  $\chi^2$ distribution for $M_y-M_x$ degrees of
freedom can be assessed. Another interesting quantity to study is the
average global correlation coefficient,
\begin{eqnarray}
\rho_{\text{avg}} &=& \frac{1}{M_x}\sum_{j=1}^{M_x} \rho_j\,.
\label{eqn:rhoavg}
\end{eqnarray}
The average global correlation coefficient can be used to tune regularisation
parameters, as discussed below. In general, one expects to find
a non-zero global correlation coefficient in the presence of non-negligible
migrations.

\subsection{Closure tests}
When using pseudo-data, generated with the help of Monte Carlo
simulations, the truth distribution $\VEC{x^{\text{truth}}}$ 
is known, so the unfolding result $\VEC{\hat{x}}$ may
be directly compared to it. Such comparisons, where pseudo-data are
unfolded and compared to the truth are often called closure tests.

The most trivial test to think of is to insert
$\VEC{\mu^{\text{truth}}}=\MAT{A}\VEC{x^{\text{truth}}}+\VEC{b}$ for the
observations $\VEC{y}$ and perform the unfolding. However, this test
is not very meaningful, because 
basically all commonly used unfolding methods will trivially result in
$\VEC{\hat{x}}=\VEC{x^{\text{truth}}}$ in this case.

More interesting closure tests are based on independent Monte Carlo
samples. For example, the $y_i$ could be drawn from Poisson distributions
given the parameters $\mu^{\text{truth}}_i$. As these 
Poisson experiments and the subsequent unfolding are repeated many
times, one gets an independent determination of the average and of the
covariance
\begin{eqnarray}
 \VEC{x^{\text{avg}}} = \langle \VEC{\hat{x}} \rangle, & \quad
 \text{and} \quad &
 (\MAT{V_{xx}^{\text{avg}}})_{jk} = \langle (\hat x_j -x_j^{\text{avg}}) (\hat x_k
   -x_k^{\text{avg}})\rangle\,,
\end{eqnarray}
where the averages are taken over the unfolded, independent Monte
Carlo samples.
The resulting $\VEC{x^{\text{avg}}}$ is expected to agree with
$\VEC{x^{\text{truth}}}$ and the resulting covariance is expected to
agree with the covariance returned by the unfolding algorithm.
This type of test, seeded from the same truth distribution as is used
to construct the matrix $\MAT{A}$ verifies the statistical properties of
the unfolding method.

The most interesting type of tests includes independent Monte Carlo
samples where the underlying truth distributions are modified.
For a good unfolding algorithm, one expects to obtain unbiased results
when unfolding observations 
drawn from the changed truth distribution using the unchanged response
matrix. For each of the independent Monte Carlo samples one can define
the $\chi^2$
\begin{eqnarray}
\chi^2_{\text{truth}}=(\VEC{\hat{x}}-\VEC{x^{\text{truth}}})^{\text{T}}\MAT{V_{xx}}^{-1}
(\VEC{\hat{x}}-\VEC{x^{\text{truth}}})\,.
\end{eqnarray}
and verify that the unfolding result is not biased. In the large
sample limit and for a completely
unbiased algorithm, the quantity
$\chi^2_{\text{truth}}$ is expected to follow a $\chi^2$ distribution
with $M_x$ degrees of freedom.

\section{Unfolding algorithms}
\subsection{Bin-by-bin correction factors}
For this method, the unfolded distribution is given by
\begin{eqnarray}
\hat{x}_i & = & (y_i-b_i)\frac{N_i^{\text{gen}}}{N_i^{\text{rec}}}\,, \\
\end{eqnarray}
where $N_i^{\text{rec}}$ ($N_i^{\text{gen}}$) is the total number of
reconstructed (generated) Monte Carlo events in bin $i$. This methods
is applicable only in the case where $M_y=M_x$ and where the bins on
detector level and truth level have a clear correspondence.

The bin-by-bin method often is used due to its simplicity;
however its results are biased significantly by the underlying Monte
Carlo distributions. The results of performing data tests using the
toy example with various unfolding methods are summarised in
Tab.~\ref{tab:results}. The use of the bin-by-bin method is clearly 
disfavoured. It yields the wrong normalisation  $Y_{\text{unf}}$ and 
$\chi^2_A$ is far from zero.
\begin{wraptable}{R}{\tablewidth}
\caption{Data tests performed on various unfolding methods.}
\label{tab:results}
\centering
\begin{tabular}{lccc}
\hline
  & $Y_{\text{unf}}$ & $\chi^2_A / N_{\text{d.f.}}$ &
$\mathrm{Prob}(\chi^2_A,N_{\text{d.f.}})$ \\
\hline
expectation & $4584$ & $N_{\text{d.f.}}$ / $N_{\text{d.f.}}$ & $0.5$ \\
\hline
bin-by-bin & $4521$ & $34.7$ / $0$ & n.a. \\
matrix inversion & $4584$ & $0$ / $0$ & n.a. \\
template fit & $4572$ & $12.0$ / $16$ & $0.74$ \\
constrained template fit & $4584$ & $12.0$ / $16$ & $0.74$ \\
Tikhonov $\tau=0.0068$ & $4584$ & $13.4$ / $16$ & $0.64$ \\
Tikhonov $\tau=0.012$ & $4584$ & $15.0$ / $16$ & $0.52$ \\
EM method $n=0$ & $4537$ & $5069$ / $0$ & n.a. \\
EM method $n=20$ & $4585$ & $5.9$ / $0$ & n.a. \\
EM method $n=100$ & $4584$ & $4.2$ / $0$ & n.a. \\
EM method $n=1000$ & $4584$ & $3.9$ / $0$ & n.a. \\
IDS $n=1$ & $4584$ & $76.8$ / $0$ & n.a. \\
IDS $n=3$ & $4584$ & $26.1$ / $0$ & n.a. \\
IDS $n=10$ & $4584$ & $8.0$ / $0$ & n.a. \\
IDS $n=30$ & $4584$ & $4.9$ / $0$ & n.a. \\
\hline
\end{tabular}
\end{wraptable}

\subsection{Matrix inversion and  template fit}
\begin{figure}[h]
\centering
\includegraphics[width=\figwidth,clip]{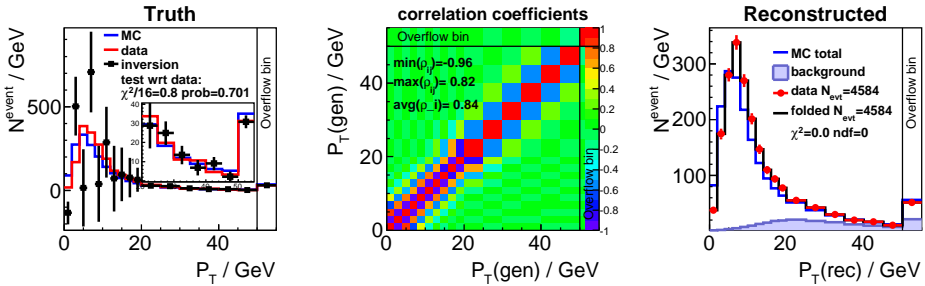}
\caption{Unfolding result using the matrix inversion method. Shown are
the unfolded distribution compared to data and Monte Carlo truth (left), the correlation coefficients
(middle) and the unfolded result folded back using the response
matrix, compared to Data and Monte Carlo (right).}
\label{fig:inversion3}
\end{figure}

Another simple unfolding method is based on inverting the
matrix $\MAT{A}$. This is possible only if the number of bins observed is
equal to the number of bins on truth level, $M_y=M_x$. The unfolded result
is given by
\begin{eqnarray}
\VEC{\hat x} &=& \MAT{A}^{-1}(\VEC{y}-\VEC{b})\,.
\end{eqnarray}
The matrix inversion returns an unbiased result, because it is a simple linear
transformation of the result and no assumptions on the probability
distributions of the $y_i$ enter the calculation. The result is
depicted in Fig.~\ref{fig:inversion3}.
While the folded-back distribution is on spot with the data and all
basic tests are fulfilled (Tab.~\ref{tab:results}), the
unfolding result shows large bin-to-bin fluctuations and
correspondingly large uncertainties and correlation coefficients
close to $-1$ for neighbouring bins. Such ``oscillation patterns'' are
often observed when solving inverse problems. The solution is
statistically correct within the uncertainty envelope given by the 
covariance matrix $\MAT{V_{xx}}$.
However, it does not correspond to a smooth curve as expected by the
physicist's prejudice on such a distribution. Extra constraints 
have to be added to the unfolding procedure in order to enforce such
behaviour. These are discussed in the next sections. In the
remaining part of this section, template fits are discussed,
corresponding to the case $M_y>M_x$. The template fits are based on a
minimisation of the expression $\chi^2_A$ (equation
\ref{eqn:chi2obs}) with respect to the $\hat{x}_j$, where the number
of degrees of freedom is $M_y-M_x>0$. Here, $M_x=17$ and $M_y=33$ are
chosen: for each truth bin two bins are used on detector level. In
addition there are overflow bins. 
Simple template fits lead to well-known biases when applied to
Poisson-distributed
data, such that the overall normalisation is not retained. For this
reason, the template fit is repeated, including a constraint on the
overall normalisation~\cite{tunfold}. The results of the template fits
with and without this constraint are included in Tab.~\ref{tab:results}.
As compared to the matrix inversion, the template fits have
somewhat reduced uncertainties and correlations, however the
large bin-to-bin fluctuations of the result are still present (not
shown in this paper). 
As summarised in Tab.~\ref{tab:results}, 
the template fits pass the data tests 
with the exception of the overall normalisation for the template fit
without constraint, which is slightly low.

\subsection{Template fit with Tikhonov regularisation}
For the constrained template fit explained above, the large
bin-to-bin fluctuations of the result can be reduced by adding an
extra term to the $\chi^2_A$ function (Eq.~\ref{eqn:chi2obs}), as
suggested by Tikhonov \cite{tikhonov}.
The function which is minimised takes the form 
\begin{eqnarray}
\chi^2_{\text{TUnfold}} =  \chi^2_A +\tau^2 \chi^2_L,
&\quad\text{where}\quad & 
\chi^2_L =
(\VEC{\hat{x}}-\VEC{x_{B}})^{\text{T}}\MAT{L}^{\text{T}}\MAT{L}(\VEC{\hat{x}}-\VEC{x_{B}})\,.
\label{eqn:tunfold}
\end{eqnarray}
The vector $\VEC{x_B}$ is the bias vector, often set to zero or to the
Monte Carlo truth. The matrix $\MAT{L}$ specifies the regularisation
conditions and is here set to the unity matrix. The
parameter $\tau$ is the regularisation strength. The case $\tau=0$
corresponds to the template fit without regularisation, whereas for
very large $\tau$ the result is strongly biased to $\VEC{x_B}$. 
Eq.~\ref{eqn:tunfold}
is modified slightly \cite{tunfold} to account for the normalisation
constraint.

\begin{figure}[h]
\centering
\includegraphics[width=\figwidth,clip]{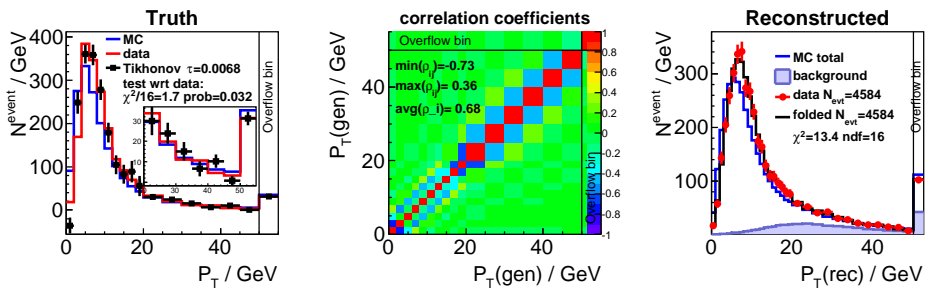}
\caption{Unfolding result using the constrained template fit with
  Tikhonov regularisation and parameter \mbox{$\tau=0.0068$}.
Further details are given in Fig.~\ref{fig:inversion3} caption.}
\label{fig:lcurveFA3}
\end{figure}

Fig.~\ref{fig:lcurveFA3} shows the unfolding result obtained for the
choice $\tau=0.0068$. As compared to the matrix inversion, the
oscillating behaviour of $\VEC{\hat{x}}$ is removed and the uncertainties
and correlations are reduced. As compared to
Fig.~\ref{fig:inversion3}, the original of the input 
distribution is visualized much better.

\subsection{Choosing regularisation parameters}
When using Tikhonov regularisation one has the 
difficulty to find an appropriate choice of the parameter
$\tau$. Similarly, the maximum number of iterations has to be chosen
in the case of iterative methods, see section
\ref{section:iterative}. Two methods are discussed in the following, the
L-curve scan, applicable for the Tikhonov case, and the minimisation
of the average global correlation coefficient, which is 
applicable to a larger class of unfolding methods.

\begin{wrapfigure}{R}{\smallhalffigwidth}
\centering
\includegraphics[width=\smallhalffigwidth,clip]{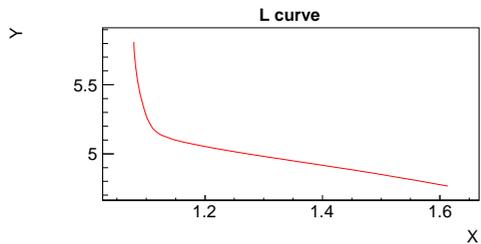}
\caption{Parametric plot of $X=\log\chi^2_A$ and
  $Y=\log\chi^2_L$ (L-curve).}
\label{fig:lcurveL1}
\end{wrapfigure}
The L-curve scan is based on investigating the variables $X:= \log\chi^2_A$ and
$Y:=\log\chi^2_L$.
The L-curve is determined by varying $\tau$ and minimising
$\chi^2_{\text{TUnfold}}$ for each choice of $\tau$. The variables 
$X$ and $Y$ are visualised on a parametric plot, as shown in
Fig.~\ref{fig:lcurveL1}.
There is a characteristic kink, i.e.~the curve is
shaped similar to the letter ``L''. The kink corresponds to the point with the largest geometric
curvature. The corresponding value of $\tau$ is chosen to set the
regularisation strength. For a review of the L-curve method, see
e.g.~\cite{hansen2000}.

\begin{figure}[h]
\centering
\includegraphics[width=\fullwidth,clip]{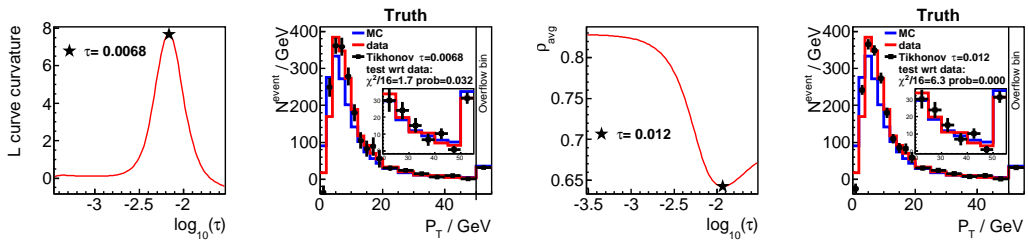}
\caption{Results of two scans of the parameter $\tau$: L-curve curvature scan
  (left panels)
and scan of average global correlation coefficients (right panels). The scans
and the corresponding unfolding result at the final choice of $\tau$
are shown.}
\label{fig:scanTau}
\end{figure}

The minimisation of the average global correlation coefficient
\cite{globalcorr} is also based on repeating the unfolding algorithm for
different choices of the regularisation parameter.
 The average global
correlation coefficient (Eq.~\ref{eqn:rhoavg}) is
recorded and the regularisation parameter is chosen at the
minimum $\rho_{\text{avg}}$. The maximisation of the L-curve curvature
and the minimisation of $\rho_{\text{avg}}$ as a function of $\tau$
are compared in Fig.~\ref{fig:scanTau}. 
In this example, but also in many other cases, the $\rho_{\text{avg}}$
minimisation yields a stronger regularisation than the L-curve
method. Both methods pass the test against data, as shown in Tab.~\ref{tab:results}.

\subsection{Iterative unfolding methods}
\label{section:iterative}
Iterative unfolding methods have been proposed since long. Here, two
such unfolding methods have been tried: the EM algorithm\footnote{the
  EM Algorithm was developed for medical image reconstruction
  by Shepp/Vardi \cite{sheppvardi} and proposed for application
  in high-energy physics by Kondor \cite{Kondor:1982ah}
  and M\"ulthei/Schorr \cite{Multhei:1986ps}.
  It was reinvented by D'Agostini
  \cite{D'Agostini:1994zf} and  is often referred to as ``iterative Bayesian
  unfolding'' in recent publications, although the similarities of
  Eq.~\ref{eqn:em} with Bayes' law are accidental \cite{sheppvardi} and
  do not ensure that this is a proper Bayesian method.} 
\cite{sheppvardi} and the IDS algorithm \cite{Malaescu:2009dm}.
The EM algorithm, in the form described in  \cite{D'Agostini:1994zf}
defines an iterative improvement of the unfolding result $x_i^{(n+1)}$,
given the result of a previous iteration, $x_j^{(n)}$,
\begin{eqnarray}
x_j^{(n+1)} = x_j^{(n)} \sum_{i=1}^{M}
\frac{A_{ij}}{\epsilon_j}\frac{y_i}{\sum_{k=1}^N A_{ik} x_k^{(n)}}\,.
\label{eqn:em}
\end{eqnarray}
In the original works  \cite{sheppvardi,Kondor:1982ah,Multhei:1986ps}, the
efficiency $\epsilon_j$ is absorbed in a redefinition  $\epsilon_j
x_j\to\tilde{x}_j$ and $A_{ij}/\epsilon_j \to\tilde{A}_{ij}$, such
that $\tilde{x}_j^{(n+1)} = \tilde{x}_j^{(n)} \sum_i
\tilde{A}_{ij} y_i / \sum_k \tilde{A}_{ik} \tilde{x}_k^{(n)}$.

The EM algorithm has two interesting properties: 
given that the start
values $x_j^{(-1)}$ and the measurements $y_i$ are all positive, the
result is bound to be positive. Furthermore, 
it converges to a maximum of the likelihood function, the likelihood
defined for the case where the measurements $y_i$ are independent and
follow Poisson distributions \cite{sheppvardi}.
However, the convergence rate can be very slow and the number of
iterations is expected to grow with the number of bins squared 
\cite{Multhei:1986ps}. While the method is expected to give unbiased results for a sufficiently large number of iterations with
Poisson distributed measurements, this is not necessarily true in other cases,
e.g.~for correlated input data. This is
evident from the fact that the covariance of the 
input data $\MAT{V_{yy}}$ does not enter Eq.~\ref{eqn:em}.

\begin{figure}[b]
\centering
\includegraphics[width=\figwidth,clip]{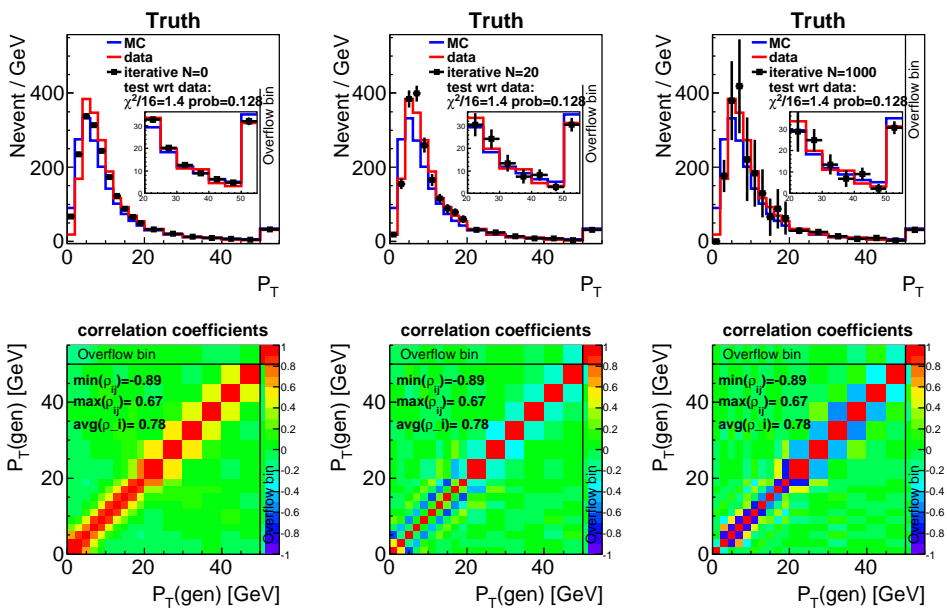}
\caption{Unfolding results using the iterative EM algorithm. Shown are
the results for a number of iterations $n=0$ (left), $n=20$ (middle) and $n=1000$ (right).}
\label{fig:iterate0201000}
\end{figure}
In high-energy physics, the EM method typically is {\em not}
iterated until it converges. Instead, the number of
iterations is fixed, and the result then still depends on the
start values $x_j^{(-1)}$. The dependence on the start values, typically
taken to be the Monte Carlo truth, provides a regularisation of the result.
Although the algorithm is expected to perform best for $M_y>M_x$, in
high-energy physics often the same number of bins is used on
detector and truth level, $M_y=M_x$. The choice $M_y=M_x$ is
also employed here. For the case of an infinite number of iterations,
the EM method with  $M_y=M_x$ is expected to agree with the results of
the matrix inversion in those cases where the $\hat{x}_j$ obtained by
the matrix inversion are all positive.

Eq.~\ref{eqn:em} has the disadvantage that background is not
included. Subtracting the background $\VEC{b}$ from
$\VEC{y}$ in the enumerator is not favourable, because it possibly spoils the
positiveness property. For this study, the denominator is
modified as follows to take into account the background,
\begin{eqnarray}
x_j^{(n+1)} & = & x_j^{(n)} \sum_{i=1}^{M}
\frac{A_{ij}}{\epsilon_j}\frac{y_i}{\sum_{k=1}^N A_{ik} x_k^{(n)}+b_i}.
\label{eqn:embgr}
\end{eqnarray}
The results obtained with the EM method for $n=\{0,20,1000\}$
iterations are shown in Fig.~\ref{fig:iterate0201000} 
and the data tests are summarised in Tab.~\ref{tab:results} for $n=\{0,20,100,1000\}$.
The results obtained for $n=0$ iterations, corresponding to the
non-iterated, so-called
{\em Bayesian unfolding} \cite{D'Agostini:1994zf} are very poor, as
visible from both the the data tests and from
Fig.~\ref{fig:iterate0201000}. All bins have positive correlations,
corresponding to a smearing rather than unfolding, and the result is
far from the truth distribution.
The data tests obtained for a low number of iterations indicate
that a certain minimum number of iterations is required to reach a proper
normalisation. The shapes of the unfolded results observed for $n=20$
and $n=1000$ iterations (Fig.~\ref{fig:iterate0201000})
are similar to the cases of Tikhonov
regularisation and matrix inversion, respectively.

\begin{wrapfigure}{R}{\verysmallhalffigwidth}
\centering
\includegraphics[width=\verysmallhalffigwidth,clip]{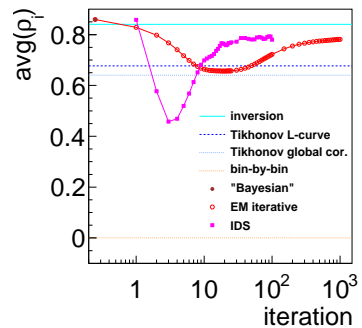}
\caption{Average global correlation coefficient as a function of the
  number of iterations.}
\label{fig:niter}
\end{wrapfigure}
Another iterative algorithm tested here is the Iterative Dynamically
Stabilised unfolding (IDS) \cite{Malaescu:2009dm}. The algorithm is
too complex to be explained here in detail. Briefly, it combines
elements of the EM iterative procedure and the bin-by-bin
unfolding, using non-linear weighting factors in each bin. For each
iteration, care is taken to preserve the data normalisation.
Because of the bin-by-bin component in the algorithm, its use is
restricted to the case $M_y=M_x$ or to cases where a clear
correspondence of bins on detector and truth level exists
\cite{Malaescu:2009dm}. The IDS algorithm is expected ultimately to
converge to the same value as the EM algorithm, however at improved
convergence speed \cite{Malaescu:private}. Results of data comparisons
after $n=\{1,3,10,30\}$ iterations are given in Tab.~\ref{tab:results}. In contrast to the EM method, in this case the data
normalisation is correctly reproduced even after only one
iteration. The observed $\chi^2_A$ indicates that a sufficiently large 
number of iterations is required to accurately match the shapes on
detector level.

\begin{wraptable}{R}{\tablewidth}
\caption{Comparison of selected unfolding results to the ``data'' truth.
  When calculating $\chi^2_{\text{truth}}$, the overflow bin is not
  included. The row labelled $\mathrm{Prob}$ corresponds to the quantile of the
  $\chi^2$ distribution for $16$ degrees for freedom.}
\label{tab:truth}
\centering
\begin{tabular}{lccccc}
\hline
  & $\chi^2_{\text{truth}} / N_{\text{d.f.}}$ &
$\mathrm{Prob}$ & $\sigma_{\text{Landau}}$ \\
\hline
MC truth & n.a. & n.a. & $2.00\phantom{{}\pm0.00}$ \\
data truth & n.a. & n.a. & $1.80\phantom{{}\pm0.00}$ \\ \hline
bin-by-bin &  $67.8$ / $16$ & $0.00$ & $2.05\pm0.05$ \\
matrix inversion & $12.6$ / $16$ & $0.70$ & $1.80\pm 0.06$ \\
constrained template fit & $13.0$ / $16$ &$0.68$ & $1.80\pm 0.06$ \\
Tikhonov $\tau=0.0068$ & $28.0$ / $16$ & $0.03$ & $1.86\pm0.06$ \\
Tikhonov $\tau=0.012$ &  $100.8$ / $16$ & $0.00$ & $1.97\pm 0.05$ \\
EM method $n=0$ & $4537$ / $16$ & $0.00$ & $2.24\pm0.02$ \\
EM method $n=20$ &  $17.9$ / $16$ & $0.33$ & $1.91\pm0.07$ \\
EM method $n=100$ & $17.3$ / $16$ & $0.37$ & $1.77\pm0.06$ \\
EM method $n=1000$ & $22.5$ / $16$ & $0.13$ & $1.75\pm0.06$ \\
IDS $n=1$ & $4573$ / $16$ & $0.00$ & $2.30\pm 0.02$ \\
IDS $n=3$ & $158.0$ / $16$ & $0.00$ & $2.27\pm 0.03$ \\
IDS $n=10$ & $20.7$ / $16$ & $0.19$ & $1.97\pm0.04$ \\
IDS $n=30$ & $13.4$ / $16$ & $0.64$ & $1.81\pm0.06$\\
\hline
\end{tabular}
\end{wraptable}
\subsection{Scan of average global correlations for iterative methods}

The dependence of the average global correlation coefficient and of
$\chi^2_{\text{truth}}/N_{D.F.}$ on the number of iterations 
is studied in Fig.~\ref{fig:niter} for the EM and the IDS algorithms.
For both algorithms, a characteristic minimum of
$\rho_{\text{avg}}$ is observed.
This is related to the fact that the
first iteration produces positively correlated results (smearing),
whereas in the limit of many iterations (matrix inversion), negative
correlation coefficients $\rho_{ij}$ appear. 
The minimum $\rho_{\text{avg}}$ is
interesting to study, because it is largely independent of the start
values and hence can be used as an objective to define the
number of iterations.
The IDS algorithm is observed to converge faster than the EM
algorithm. The minimum in
$\rho_{\text{avg}}$ is reached after $3$ ($20$) 
iterations for the IDS (EM) method. The minimum value of $\rho_{\text{avg}}$
determined for the EM method is similar to the minimum observed in
template fits with Tikhonov regularisation, whereas the IDS minimum
is significantly lower. Most probably this is related to the fact that
the IDS algorithm has a bin-by-bin correction component (bin-by-bin
trivially results in $\rho_{\text{avg}}=0$).

\subsection{Comparisons on truth level}

%
To assess the quality of the unfolding in greater detail, closure
tests are performed. The unfolded data are compared to the data truth,
which is known for the toy example. In addition, fits of the original
Landau function are performed to the unfolded distributions. The
comparisons to truth are summarised in Tab.~\ref{tab:truth}.
Here, the comparisons are based on unfolding the ``data'' and
comparing to the truth. For more detailed tests, toy studies using the
``data'' truth would have to be performed in order to assess the quality of
the expectation values $\langle\hat{x}\rangle$ and the distribution of
$\chi^2_{\text{truth}}$.

The bin-by-bin method, the Tikhonov method with large $\tau$ and the
iterative methods with small number of iterations all result in
unacceptable 
biases of the extracted width $\sigma_{\text{Landau}}$. As expected, the matrix
inversion and constrained template fit results are not biased. The
Tikhonov method with L-curve scan, resulting $\tau=0.0068$, gives
acceptable results. The EM method works well for a sufficiently large
number of iterations, $n\gtrsim20$, where $n=20$ was determined in the
scan of $\rho_{\text{avg}}$. The IDS method also performs
well for a sufficiently large number of iterations $n$; however 
$n=3$ determined in the scan of $\rho_{\text{avg}}$ does not give a
satisfactory result.

\section{Summary and Conclusions}

A selection of methods to unfold binned distributions is studied:
bin-by-bin correction factors,
matrix inversion, template fits with Tikhonov regularisation, 
iterative methods. The bin-by-bin methods leads to biased results and
should not be used. Matrix inversion and constrained template fits
give unbiased results. However, the result typically suffer from large
bin-to-bin correlations, large uncertainties and bin-to-bin oscillation patterns.

The oscillations are reduced in the other unfolding methods using
regularisation techniques. These damp the fluctuations and reduce 
bin-to-bin correlations, at the cost of introducing biases.
There are free parameters which have to be tuned to obtain a good
compromise between bias and damping.

Template fits with Tikhonov regularisation seem to give good results
when choosing the regularisation parameter by means of the L-curve
method. For the iterative EM method, an interesting choice is the
minimisation of the average global correlation coefficients. The IDS
iterative method also has been tested but seems to require a
different objective to optimise the number of iterations.

In general, methods with Tikhonov regularisation have the advantage,
that there is a natural transition to unbiased results, by setting
the $\tau$ parameter to zero. In contrast, the iterative methods start
from a fully biased results and the number of iterations required to
reach the unbiased result is {\em a priory} unknown. Furthermore, in
the case of bin-to-bin correlated or non-Poisson distributed
measurements, it is not clear whether the iterative methods 
converge to an unbiased estimator.

In summary, no matter which unfolding method is used, detailed closure tests
are required to quantify the level of bias introduced by the unfolding.

\begin{flushleft}

\end{flushleft}

\end{document}